# Package Hermeticity Testing With Thermal Transient Measurements

András Vass-Várnai[1,2], Márta Rencz[1,2]
vassv@eet.bme.hu, rencz@micred.com
[1]Budapest University of Technology and Economics (BUTE), Dept. of Electron Devices
Budapest, Hungary
[2]MicReD Ltd
Budapest, Hungary

The rapid incursion of new technologies such as MEMS and smart sensor device manufacturing requires new tailor-made packaging designs. In many applications these devices are exposed to humid environments. Since the penetration of moisture into the package may result in internal corrosion or shift of the operating parameters, the reliability testing of hermetically sealed packages has become a crucial question in the semiconductor industry. Thermal transient testing, a well known technique for thermal characterization of IC packages [1] can be a suitable method for detecting hermeticity failures in packaged semiconductor and MEMS devices. In the paper this measuring technique is evaluated. Experiments were done on different measurement setups at different environment temperature and RH levels. Based on the results, a new method for package hermeticity testing is proposed.

## I. INTRODUCTION

It is a well known fact that the relative humidity (RH) of the ambient atmosphere has a very low effect on its thermal properties such as thermal conductivity and thermal capacitance, especially at room temperatures. At elevated temperatures this effect gains higher importance, but it is still hard to detect the small changes of the mentioned parameters with thermal transient measurements.

A possible solution could be based on the fact that water has a good thermal conductivity coefficient (0.58W/m·K), compared to the air (0.0257 W/m·K) and a relatively high specific heat of evaporation ($2,25 \times 10^6$ J/kg). An encapsulated porous adsorbent material in the package under test may serve as a sensing layer. The humidity content of the porous layer, the rate of adsorption and desorption is highly dependent on the relative humidity inside the package. As the RH increases the number of the adsorbed molecules should increase and vice versa [2]. As the water fills the pores the overall thermal conductivity of the layer increases [3]. In case of high surface adsorbents the decrease of the thermal resistance can be significant.

## II. EXPERIMENTAL

### A. Measurement on open packages

BC 300 transistors mounted on TO-39 type packages were used to evaluate the effects of the moisture content of the air at different RH levels. The metal caps were removed from the packages in order to expose the transistors to the environment. Thermal transient measurements were performed with the *T3Ster* equipment [4] to measure the thermal resistance of the structure between the junction and the ambient at two different RH levels. Precise RH environments were set over two different saturated salt solutions [5]. LiCl and $KNO_3$ solutions were used resulting 11.3% and 93.58% RH at 25°C. Cooling curves were taken up, 0.1W power step was applied.

### B. Application of porous materals

In order to prove the applicability of high surface porous materials for package hermeticity testing, model experiments were carried out. A 0.1 mm wide gypsum layer was placed between a BD245C power NPN transistor in a SOT-93 package and an aluminum heat-sink. This type of transistor is designed to have a relatively small junction to case thermal resistance, 1.56 K/W. Prior to the measurements the gypsum was dried for 48 hours.

The transistor was used as a heat source in this measurement setup. The heat was forced through the gypsum layer into the heat-sink. Thermal transient measurements were made with the *T3Ster* equipment to measure the thermal resistance of the gypsum at different moisture contents. The first measurement was made on the completely dry material, after that it was wetted with a small drop of water. The same thermal resistance measurements were performed one after the other on the wet system.

## III. RESULTS AND DISCUSSION

The goal of both experimental setups was to identify the changes of the humidity content of an environment with thermal transient measurements. This method is suitable and widely used in the industry for the thermal characterization of semiconductor packages. The semiconductor device inside the package is heated up (or cooled down) by a power step. The temperature change is measured through corresponding temperature sensitive electrical parameters, in our case the emitter-base voltage ($U_{EB}$) of the transistors. The measurement of the $U_{EB}$ change begins in 1µs after the power step and it is continuous until the voltage reaches a steady state. Knowing the temperature sensitivity of the device, the voltage function can be easily transformed into a temperature transient curve.

The resulting transient curve is a unit-step response function, it is characteristic to the structure inside the package and its environment. Applying the NID method (Network Identification with Deconvolution), from the transient curve a Cauer type thermal impedance model can be calculated, which is a ladder-like model of the heath flow path consisting of thermal resistances and thermal capacitances [6]. From the Cauer model structure functions are generated. The cumulative structure function provides a map of the cumulative thermal capacitances with respect to the thermal resistances measured from the location of the heating to the ambient [7]. The changes in the cumulative structure function indicate reaching new materials in the heat flow path, and their distance on the horizontal axis gives the partial thermal resistances between them, as it can be seen in Fig. 1.

*A. Junction to ambient thermal measurements*

In case of silicon devices exposed to the environment, we expected that the variation of the thermal resistance of the air as a function of its humidity content can be measured. The junction to ambient thermal resistance can be easily calculated if the measured temperature rise is divided by the applied power step:

$$R_{thja} = \Delta T_j / P \qquad (1)$$

The results of the measurements have proven that the humidity content of the air has a very low effect on its thermal conductivity coefficient. Applying the same power-step on the device at the two different RH levels, 11.3% and 93.58%, the same temperature elevation was measured. In addition to this it takes a very long time, a few minutes or even more, depending on the structure of the device to reach the thermal equilibrium which is essential to obtain valid junction-to-ambient thermal resistance values. This method is not applicable for package hermeticity testing in this form.

*B. Application of porous materials*

As the effect of the RH changes of the air on its thermal conductivity is too small to be measured with thermal transient measurements, a methodology had to be found to enhance it. The application of high surface porous materials is a possible solution. The thermal conductivity of porous materials is highly affected by their porosity. If we assume dry environment and 0% RH, the pores and cavities inside their structure contain only dry air. As the air has a low thermal conductivity the overall thermal resistance of the structure will be large. At low RH levels the water molecules from the gas phase start to adsorb on the wall of the pores and cavities. As the RH increases more and more layers of water adsorbs on the pore walls until at a certain RH level capillary condensation occurs [8]. As the water is a more than 20 times better heat conductor than the air, the overall thermal resistance of the layer decreases with the increasing humidity content. Naturally the moisture can penetrate into the pores from liquid phase as well, causing the same thermal effect.

Fig. 2. shows the heat flow path from the heated junction through the inner features of the SOT-93 package and the gypsum layer at different moisture contents into the heat-sink. The measured structure shows the highest junction-to-ambient thermal resistance (measured between the origin and the singularity point where the function goes to the infinite thermal capacitance values) when it is completely dry. The lowest junction-to-ambient thermal resistance value is shown when the structure has the highest moisture content, right after wetting it with a water droplet. The structure functions between the two states were measured one after the other with app. 2 minutes differences. As the humidity evaporates from the structure, the partial thermal resistance of the gypsum layer increases. The diverging point of the cumulative structure functions gives a good estimation for the junction-to-case thermal resistance of the package, in other words until that point the heat propagates inside the package, after that point the heat propagates through the gypsum layer. Assuming -2mV/K for the sensitivity of the silicon device, this point is very close to the 1.56 K/W value obtained from the datasheet of the transistor.

Fig. 3. shows the same functions, but in a linear view along the capacitance axis. In this view the small decrease of the thermal capacitance of the gypsum layer can be identified, as the humidity evaporates from the porous structure. This effect is very slight compared to the variation of the thermal resistances with the humidity content.

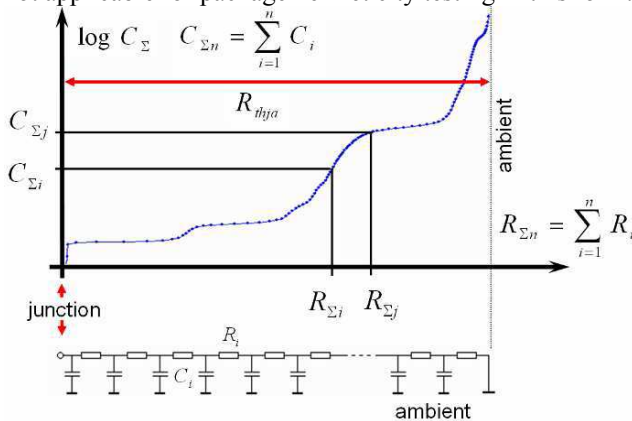

Fig. 1. Concept of the cumulative structure function.

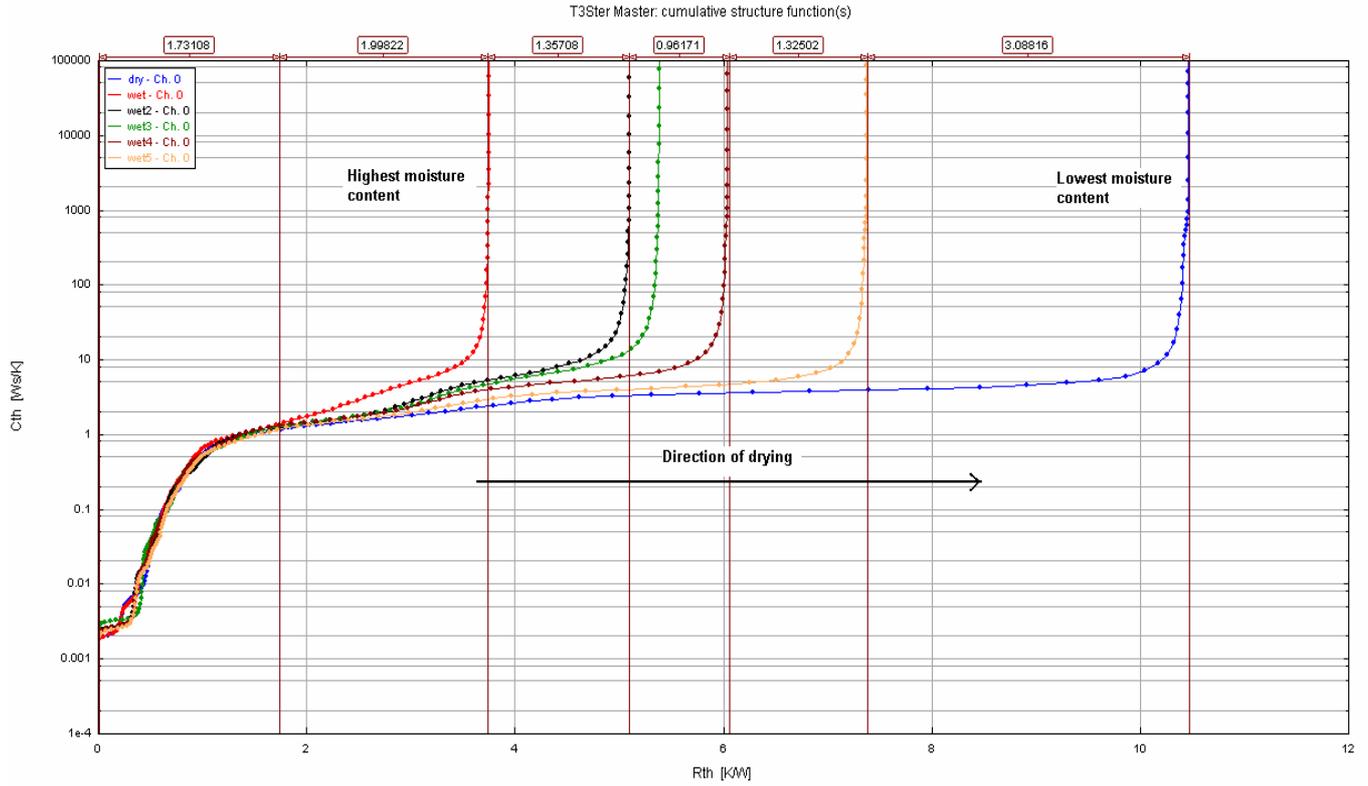

Fig. 2. Cumulative structure functions indicating the decreasing moisture content of the gypsum layer

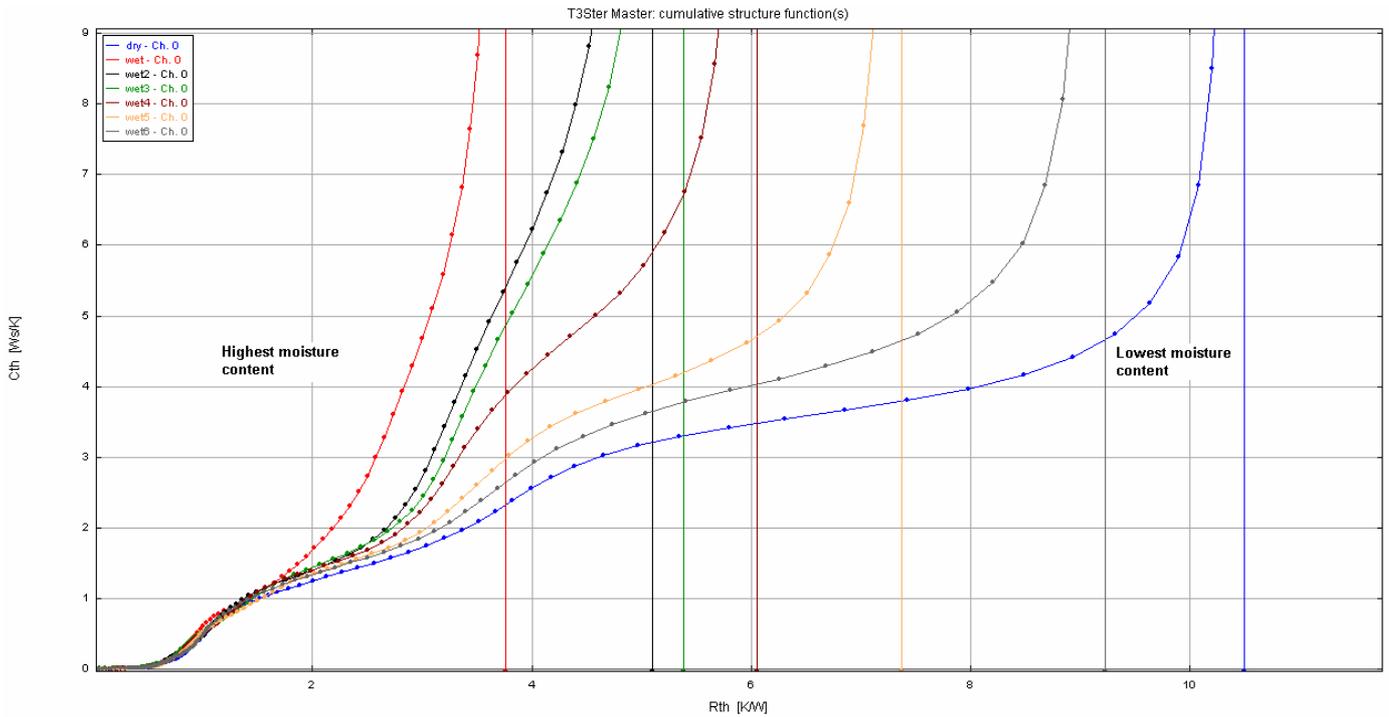

Fig. 3. Linear view of the cumulative structure functions to visualize the change of the thermal capacitances

III. CONCLUSIONS

In this paper we have shown that the thermal transient testing is not an applicable method to evaluate the hermeticity of IC packages, as the change of the thermal conductivity of the air as a function of the relative humidity is negligible. However with the application of high surface porous materials the humidity changes of the environment can be indicated. We have observed significant increase of the thermal resistance of the gypsum layer as it lost its moisture content.

The promising results of the model experiment encourages us to design and develop a microstructure which is capable of indicating the moisture content of its environment and therefore it can be used for package hermeticity testing as well. A porous alumina layer [9] will be used for adsorbing the moisture from its environment, and a simple p-n junction will be used as a heater and sensor.

We plan to find other applicabilies of these results, as it is very important both in the food and the building industry to assess the drying of porous materials.

ACKNOWLEDGMENT

This work was supported by the PATENT IST-2002-507255 Project and the NANOPACK FW7 No. 216176/2007 IP Project of the EU.